\documentclass[10pt,aps,amsmath,amssymb,nccmth,prb,twocolumn,notitlepage,showpacs,superscriptaddress]{revtex4-2}
\usepackage{amsmath, nccmath}
\usepackage{amssymb}
\usepackage{graphicx}
\usepackage{epstopdf}
\usepackage[colorlinks=true]{hyperref}
\usepackage{physics}
\usepackage{mathrsfs}
\usepackage{comment}

\renewcommand\({\begin{equation}}	% quick macro for equation with numbers
\renewcommand\){\end{equation}}
\renewcommand\[{\begin{eqnarray}}	% quick macro for equation with numbers
\renewcommand\]{\end{eqnarray}}

\usepackage[dvipsnames]{xcolor}

\begin{document}

\title{Perspective: Quantum Computing on Magnetic Racetrack}

\author{Ji Zou}
\affiliation{Physics Department, King Fahd University of Petroleum and Minerals, 31261, Dhahran, Saudi Arabia}
\affiliation{Quantum Center, KFUPM, Dhahran, Saudi Arabia}
\author{Jelena Klinovaja}
\affiliation{Department of Physics, University of Basel, Klingelbergstrasse 82, 4056 Basel, Switzerland}
\author{Daniel Loss}
\affiliation{Physics Department, King Fahd University of Petroleum and Minerals, 31261, Dhahran, Saudi Arabia}
\affiliation{Quantum Center, KFUPM, Dhahran, Saudi Arabia}
\affiliation{RDIA Chair in Quantum Computing}
\affiliation{Department of Physics, University of Basel, Klingelbergstrasse 82, 4056 Basel, Switzerland}

\begin{abstract}
Magnetic domain walls have long been pursued as carriers of classical information for storage and processing. With the ability to create, control, and probe domain walls at the nanoscale, they are recently recognized as an ideal platform for studying macroscopic quantum effects and provide a natural blueprint for building scalable quantum computing architectures. In particular, the experimentally demonstrated high mobility of domain walls makes them not only suitable as stationary qubits but also as flying qubits, which may offer  advantages over currently explored quantum computing platforms. In this Perspective, we outline our current understanding of the essential ingredients and key requirements for realizing universal quantum computation based on magnetic domain walls. We highlight promising concrete material platforms and identify the experiments that are still needed to advance this concept. We also discuss the potential challenges and point to new opportunities in this emerging research direction at the interface between magnetism and quantum information science.
\end{abstract}

\date{\today}
\maketitle

\section{Introduction}
Large scale quantum computers promise to solve certain problems far beyond the reach of classical machines~\cite{shor_1994}. This has motivated an intense search for scalable physical platforms, pursued in parallel by researchers from diverse branches of physics. A wide range of candidates is being actively explored, including spin based qubits in quantum dots~\cite{PhysRevA.57.120, Basso2019prl, Qiao:2020ncom,Hendrickx:2021tv,Philips:2022ur,petta2022sa,zou2024spatially,bosco2024high}, trapped ions~\cite{Blinov2004nature, volz2006prl, Blatt2008nature, Bruzewicz2019aip}, and superconducting circuits~\cite{Koch_pra_2007, Barends_prl_2013,Wendin_2017}.  Each approach relies on distinct underlying physics and therefore offers their unique advantages while also facing specific challenges. At present, no single route has emerged as definitively superior or fully satisfies the combined demands of coherence, controllability, connectivity, and scalability. In this regard, new qubit realizations may provide fresh opportunities, introduce complementary capabilities, expand the design space for quantum architectures, and also  offer fundamentally different advantages that existing platforms cannot easily provide. These possibilities have fueled sustained interest in alternative qubit encodings and new physical systems capable of supporting coherent quantum states and enabling their precise control and probing, with potential for universal quantum computation.

Among the various possibilities, magnetic materials form a particularly intriguing class. From  early permanent magnets to modern hard disk drives and magnetic random access memory,  magnetism has long been a backbone of information storage and processing in human history. Building on this evolution, the concept of racetrack memory proposed in Ref.~\cite{Parkin190} introduced a new paradigm in which trains of domain walls are driven along magnetic nanowires by spin-polarized currents. It offers the prospect of a high-density, low-cost, and nonvolatile storage technology, and has inspired extensive experimental and theoretical efforts over the past decades~\cite{Ryu:2013wh, Yang2015uw, Kim2017natmat,Yang:2021wy,Guan:2021vo, Blasing:2018vb,Yoshimura:2016uk,Zangprl,jiprl2020,Zang2017prb, jivortex, quantumvortex,Yaroslavreview,Kumar:2022vy,doi:10.1063/5.0042917,Grollier:2020ur,PhysRevLett.125.207202,PhysRevLett.121.127701,Daltonenergy,parkin2015memory}.  Crucially, this body of work demonstrated that domain walls can be created, pinned, and transported with high precision along nanoscale wires, establishing an engineering foundation
that a quantum extension of the racetrack concept can directly inherit.

The Landau-Lifshitz-Gilbert equation and micromagnetic simulations have been remarkably successful at describing
domain wall dynamics, reinforcing the long-standing view of magnetic textures as classical objects. This picture, however, is expected to break down when domain walls are confined to the nanoscale and cooled to millikelvin
temperatures, where quantum fluctuations of the collective coordinates can no longer be
neglected~\cite{daniel2016prb,daniel_1997_prb,loss1992suppression}. The recent emergence of two-dimensional van der Waals magnets, with atomically thin layers, strong anisotropy, and highly tunable interactions, has enabled precise control, imaging, and manipulation of domain walls at precisely the length scales where quantum effects are expected to become important~\cite{huang2017layer,Song_science_2021,ziebel2024crsbr,zur2023magnetic,tschudin2024imaging}.
This rapid experimental progress has  sparked growing interest in quantum properties of magnetic textures~\cite{takei2017spin,Yqwinding,PhysRevB.97.064401,petrovic2025colloquium} and naturally raises the question of whether such topological textures can be pushed into the quantum regime and harnessed as new building blocks for quantum computation.

In this Perspective, we address this question by examining the theoretical foundations, material requirements, and
experimental path toward quantum computation with domain wall qubits. We first show how the chirality of a magnetic domain wall provides a natural qubit degree of freedom, analyzed through both a semiclassical collective-coordinate framework and density matrix renormalization group (DMRG) simulations in the fully quantum spin-1/2 limit. We then discuss the essential ingredients for universal quantum computation, including single-qubit gates, two-qubit entangling operations, initialization, and readout, and highlight how the intrinsic mobility of domain walls enables a flying qubit concept in which quantum information is physically transported along the racetrack, a capability unique among solid-state platforms. We identify promising material candidates, with a detailed quantitative assessment of the van der Waals magnet CrSBr, and lay out a concrete experimental roadmap.
Finally, we look toward broader possibilities, including other topological textures and hybrid architectures that combine domain wall qubits with existing quantum platforms to make use of their complementary strengths.

\begin{figure}[!t]
\centering
\includegraphics[width=1\columnwidth]{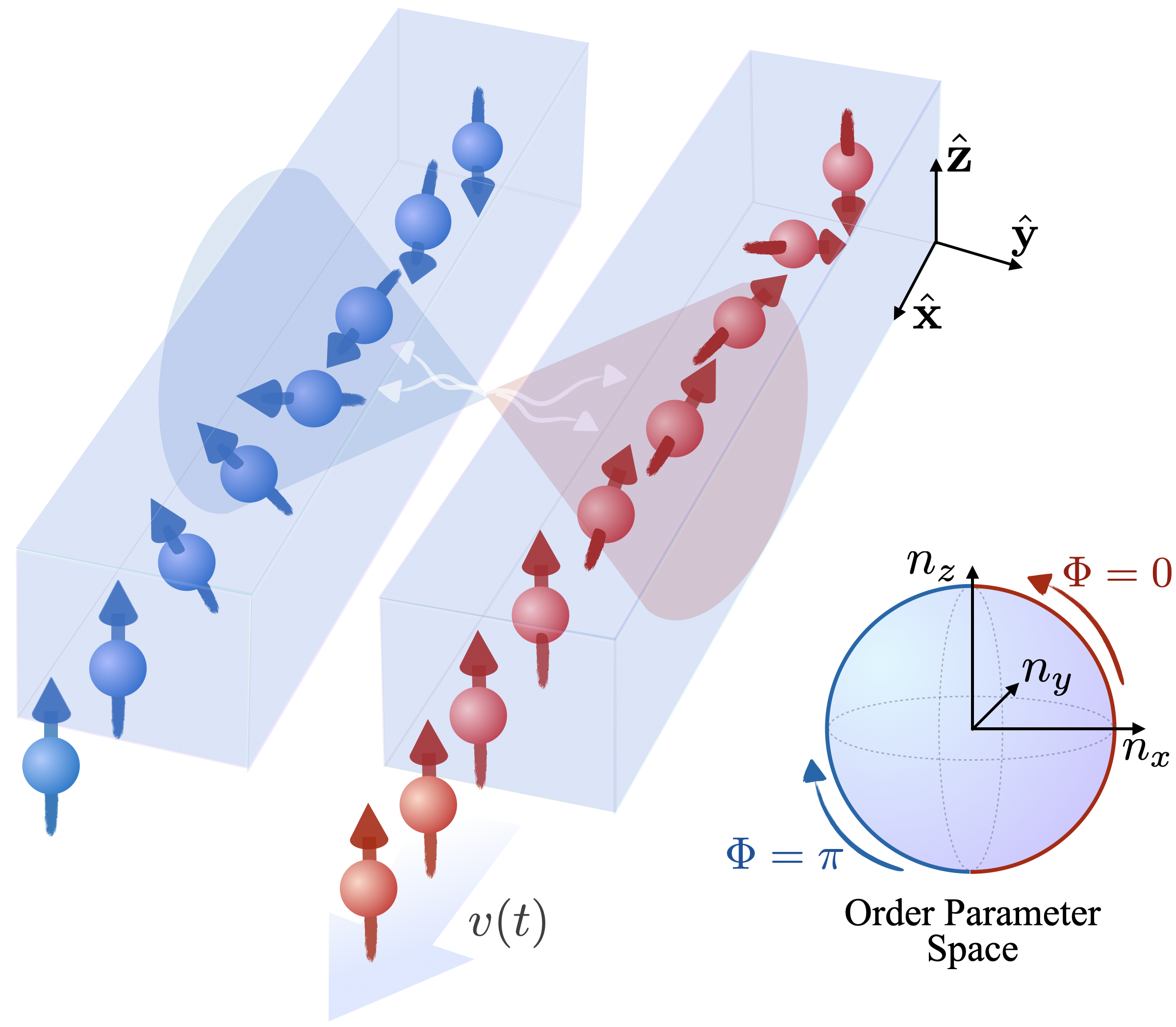}
\caption{\textbf{Domain wall qubit on magnetic racetrack.}
Two N\'{e}el-type domain walls with opposite chiralities are
illustrated on a nanowire with easy axis along $\hat{z}$. The red
($\Phi = 0$) and blue ($\Phi = \pi$) configurations correspond to
positive and negative chirality, respectively, with the
magnetization rotating through the wall in opposite senses. These
two chiralities trace distinct paths on the order parameter sphere
(lower right) and serve as the two basis states of the domain wall
qubit. A current-driven velocity $v(t)$ allows the domain wall to be
transported along the nanowire, carrying its internal quantum state
with it and thereby functioning as a flying qubit. }
\label{f1}
\end{figure}

\section{Domain Walls as Quantum Two Level Systems}\label{sec:II}
\subsection{Quantized domain wall modes in confined geometries}
Magnetic domain walls are quasi-one-dimensional topological solitons that occur widely in magnetic materials~\cite{kim2023mechanics}. Their structure and dynamics have been studied for decades, including how they can be driven, controlled, and stabilized. Such investigations span ferromagnets, ferrimagnets, and antiferromagnets, and more recently extend to monolayer or few layer two dimensional van der Waals magnets such as Fe$_3$GeTe$_2$, CrI$_3$, and CrSBr~\cite{deng2018gate,yang2022magnetic,guan2025highly,thiel2019probing}.  These systems provide versatile platforms for exploring both classical and emergent quantum properties of domain walls.

\begin{figure}[!t]
\centering
\includegraphics[width=\columnwidth]{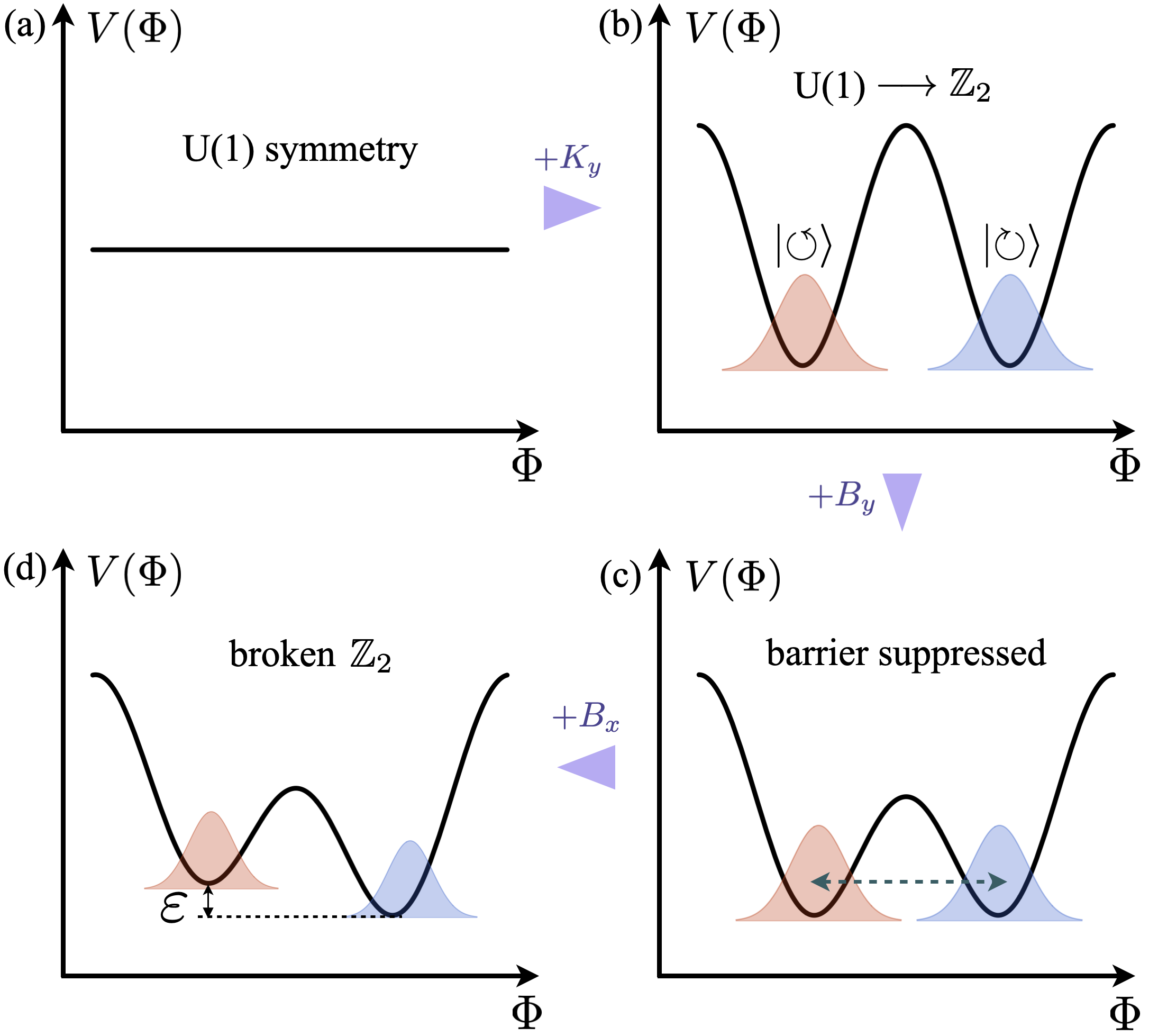}
\caption{ \textbf{Engineering a  qubit from the domain wall
collective coordinate $\Phi$.} (a)~In a uniaxial magnet with only
exchange $J$ and easy-axis anisotropy $K_z$, the potential
$V(\Phi)$ is flat, reflecting the full $U(1)$ rotational symmetry
of the domain wall angle. (b)~Introducing a hard-axis anisotropy
$K_y$ breaks $U(1) \to \mathbb{Z}_2$, generating a periodic
double-well potential whose minima correspond to the two chirality
states $\ket{\circlearrowleft}$ ($\Phi = 0$) and
$\ket{\circlearrowright}$ ($\Phi = \pi$). The wells are
separated by a large barrier, and tunneling between them is
exponentially suppressed. (c)~Applying a magnetic field $B_y$
along the hard axis tilts the spin texture out of the easy
$xz$-plane, suppressing the potential barrier and enabling a
sizable tunneling rate between the two chirality states (dashed
arrows). (d)~An additional field component $B_x$ breaks the
residual $\mathbb{Z}_2$ symmetry, introducing a tunable energy
detuning $\varepsilon$ between the two wells.}
\label{f2}
\end{figure}

To describe the domain wall structure, we consider a magnetic texture that varies along the $x$ direction, as illustrated in Fig.~\ref{f1}. In a simple uniaxial magnet with its easy axis oriented along $z$, the domain wall is stabilized by the competition between the exchange interaction $J$ and the anisotropy energy $K_z$. The resulting profile of the order parameter field $\vb n(x)$, which captures the low-energy magnetization dynamics, takes the following form,
\(
n_x + i n_y = e^{i\Phi} \, \sech\!\left({x - X}\right), \;\;\;
n_z = \tanh\!\left({x - X}\right).
\)
Here and throughout, distances are measured in units of the domain wall width $\lambda = Sa\sqrt{J/K_z}$, where $a$ is the lattice constant and $S$ is the spin length.
 Importantly, the wall possesses two zero modes: the collective coordinate $X$, which specifies the position of the wall in real space, and the angle $\Phi$, which describes the azimuthal orientation of the domain wall in spin space. Figure~\ref{f1} shows two Néel-type walls with $\Phi = 0$ (red) and $\Phi = \pi$ (blue). In a system with a single easy axis, $\Phi$ is a free parameter and all values from $0$ to $2\pi$ correspond to energetically equivalent domain wall configurations as shown in Fig.~\ref{f2}(a).

When domain walls are confined to nanometer scales, as occurs in materials with strong magnetic anisotropy, their dynamics are expected to become quantum mechanical at sufficiently low temperatures. In this regime, the collective coordinate $\Phi$ can no longer be treated as a classical variable and must instead be quantized, opening the possibility of using this degree of freedom to encode quantum information. A natural choice is to identify domain wall chirality as the qubit degree of freedom, with the positive-chirality configuration $\ket{\circlearrowleft}$ and the negative-chirality configuration $\ket{\circlearrowright}$ serving as the two logical states~\cite{Zou2023prr}. Realizing such a qubit requires engineering the magnetic system so that the effective potential $V(\Phi)$ develops a double-well form whose minima correspond to these chiral configurations. Quantum tunneling between the wells then produces coherent superpositions, allowing domain wall chirality to function as a natural qubit. This idea applies to various classes of magnetic materials. Here we focus on ferrimagnets, which offer potentially higher operation speeds than ferromagnets~\cite{trif2026cavity} while remaining easier to control and detect than antiferromagnets~\cite{PhysRevB.97.064401}.

For a domain wall pinned at a fixed position, the spatial coordinate $X$ is effectively frozen, and the low-energy dynamics reduce to those of the collective angle $\Phi$. To shape the potential $V(\Phi)$ into a double-well profile suitable for qubit encoding, two additional ingredients beyond the exchange coupling $J$ and the easy-axis anisotropy $K_z$ are required. First, a hard-axis anisotropy $K_y$ penalizes spin orientations along $y$,
breaking the $U(1)$ rotational symmetry of $\Phi$ down to $\mathbb{Z}_2$ and selecting two preferred chirality configurations as the potential minima, as shown in the Fig.~\ref{f2}(b). Second, an external magnetic field $\vb{B}$ in the $xy$-plane provides two independent control knobs: its $y$-component physically tilts the spin texture out of the easy $xz$-plane, effectively suppressing the potential barrier between the two chirality states and yielding a sizable tunneling rate between them as illustrated in Fig.~\ref{f2}(c), while its $x$-component breaks the residual
$\mathbb{Z}_2$ symmetry and introduces a tunable detuning between the two chirality states [see Fig.~\ref{f2}(d)].

Together, these ingredients give the domain wall the effective
Lagrangian
\begin{equation}
L(\Phi) = \frac{M}{2}\dot{\Phi}^2 - V(\Phi), \label{eq:potential}
\end{equation}
with  $V(\Phi) = 2NK_y(\sin^2\Phi - 2b_x \cos\Phi -
2b_y\sin\Phi)$. Here $M = N\hbar^2/2J$ is the effective mass, with
$N$ the number of spins within the domain wall, and $b_i = \pi
g\mu_B S_e B_i / 4K_y$ ($i = x, y$) is the dimensionless magnetic
field measured in units of the hard-axis anisotropy, with $S_e$
denoting the excess spin per site in the ferrimagnet. The relevant
energy hierarchy is $g\mu_B S_e B, K_y < K_z < J$, ensuring
that the domain wall remains well defined as a topological object
while its internal degree of freedom $\Phi$ is governed by the
softer scales $K_y$ and $B$.

Within the two lowest-lying states of this engineered double-well
potential, the chirality degree of freedom naturally forms a
tunable two-level system  described by the effective Hamiltonian~\cite{Zou2023prr}
\begin{equation}
H = \frac{\varepsilon}{2}\sigma_z
  - \frac{t_g}{2}\sigma_x, \label{eq:qubit_H}
\end{equation}
 in basis $\{ \ket{\circlearrowleft},  \ket{\circlearrowright} \}$ where the detuning $\varepsilon = -8NK_y b_x$ is controlled by the
magnetic field along $x$, and the tunneling amplitude
\begin{equation}
t_g \approx 4\hbar\omega_0
\sqrt{\frac{S_0}{2\pi\hbar}}\,
\exp\!\left(-\frac{S_0}{\hbar}\right) \label{eq:tg}
\end{equation}
is governed by the field along $y$, which sets the barrier height $V_0 = 2NK_y(1 - b_y)^2$.
Here $S_0 \approx 4V_0/\omega_0$ is the instanton
action, with
the level spacing $\hbar\omega_0 = 2\sqrt{2JK_y(1 - b_y^2)}$.
This mapping onto a standard qubit Hamiltonian, with independently
tunable $\varepsilon$ and $t_g$, establishes the domain wall
chirality as a fully controllable quantum two-level system.

\subsection{Flying domain wall qubit on racetrack}
A central challenge in scaling quantum processors is connectivity: how to communicate qubits that are not nearest neighbors~\cite{vandersypen2017interfacing}. In most solid-state platforms, qubits remain fixed at predefined locations on a chip, and any long-range interaction must be mediated by additional components~\cite{mi2018coherent,landig2018coherent,nigg2017superconducting,Daniel2012prx,Daniel2013prx,PhysRevB.101.014416,zou2022prb,xue2025directional,driessen2025robust}, inevitably complicating the architecture. This limitation is not merely an engineering inconvenience;  it also fundamentally constrains which quantum
error-correcting codes can be efficiently implemented on a given platform. For example, recent progress in low-density parity-check codes has shown that better encoding rates and fault-tolerance thresholds than the surface code are achievable, but only at the cost of demanding nonlocal stabilizer checks spanning distant qubits~\cite{panteleev2021quantum,panteleev2021degenerate,PRXQuantum.2.040101}. On hardware restricted to nearest-neighbor connectivity, the routing overhead required to realize such operations can erode much of this  advantage.

Flying qubits offer a conceptually different path forward. Rather than engineering ever more complex coupling schemes between static qubits, the quantum information itself is physically transported to where it is needed. This is precisely where magnetic domain walls become compelling. On a racetrack, qubit transport is not an added capability attached to a stationary qubit; it is an intrinsic property of the platform, inherited from the classical racetrack memory concept that was designed for exactly this purpose~\cite{Parkin190}. A domain wall naturally combines information storage, transport, and processing within a single mobile object.

When the nanoscale domain wall is set into motion, the spatial coordinate $X$ must also be treated quantum mechanically, and its interplay with the chirality degree of freedom $\Phi$ endows the domain wall qubit with qualitatively new features. The Hamiltonian governing the two soft modes couples the translational motion of the wall to its internal chirality through a $\Phi$-dependent gauge field originating from the spin Berry phase intrinsic to magnetic systems. Upon projecting $\Phi$ onto the chirality qubit subspace, one obtains~\cite{Zou2023prr}
\begin{equation}
H = \frac{\varepsilon}{2}\sigma_z
  - \frac{t_g}{2}\sigma_x
  + \frac{(\hat{P}
    + \beta\,\sigma_z)^2}{2M}
  + \frac{M\omega_p^2}{2}[\hat{X} - X_0(t)]^2,
\end{equation}
where $[\hat{X}, \hat{P}] = i\hbar$. The domain wall is confined by a harmonic potential of frequency $\omega_p$, centered at $X_0(t)$, which can be precisely controlled by magnetic or electric fields or by spin-polarized currents~\cite{Ryu:2013wh, Yang2015uw, Kim2017natmat,Yang:2021wy,Guan:2021vo, Blasing:2018vb,Yoshimura:2016uk}. The third term reveals that the translational momentum of the domain wall couples directly to
the qubit pseudospin through an effective spin-orbit interaction. The coupling strength is set by $\beta \approx \pi N\hbar(S_e + g\mu_B B_y/4J)$, which defines a spin-orbit length $l_{\text{so}} = \hbar/\beta$: the distance over which translational motion induces a full qubit rotation.

Switching to the moving frame of the domain wall and projecting onto the orbital ground state reduces the flying qubit Hamiltonian to a compact form,
\begin{equation}
\mathcal{H}_m = \frac{\tilde{\varepsilon}(t)}{2}\,\sigma_z
              - \frac{\tilde{t}_g}{2}\,\sigma_x, 
\end{equation}
with renormalized tunneling and detuning given by
\begin{equation}
\tilde{t}_g = t_g\,\exp(-l_p^2/l_{\text{so}}^2), \;\;\;
\tilde{\varepsilon}(t) = \varepsilon
  + \frac{2\hbar\,\dot{X}_0(t)}{l_{\text{so}}}, \label{parameter}
\end{equation}
where $l_p = \sqrt{\hbar/M\omega_p}$ is the characteristic length of the confining potential, reflecting the positional uncertainty of the domain wall. We note that the spin-orbit interaction suppresses the tunneling
amplitude by an exponential factor controlled by the ratio $l_p/l_{\text{so}}$: the larger the positional spread of the wall relative to the spin-orbit length, the stronger the suppression. The physical origin of this suppression has a close analog in semiconductor physics: the $g$-factor renormalization of a quantum dot with spin-orbit coupling~\cite{froning2021strong,dmytruk2018renormalization}. In the qubit pseudospin language, the tunneling term $-(t_g/2)\,\sigma_x$ acts as an effective magnetic field along the $x$-direction. In the frame that removes the spin-orbit gauge field, this transverse field rotates around the $z$-axis as a function of position along the racetrack, completing one full rotation over a distance $\pi l_{\text{so}}$. The quantum zero-point motion of the confined wall samples this rotating field over a range $\sim l_p$: when $l_p \gg l_{\text{so}}$, the wall averages over many rotation periods and the transverse field cancels out, exponentially suppressing the tunneling; when $l_p \ll l_{\text{so}}$, the wall remains well localized within a single rotation period and the tunneling is essentially unaffected.

More strikingly, the motion generates a velocity-dependent contribution to the detuning. {This contribution takes the  form of Eq.~\eqref{parameter}, $\tilde{\varepsilon}(t) = \varepsilon + 2\hbar v/l_{\text{so}}$, proportional to the externally controlled wall velocity $v = \dot{X}_0(t)$, with the detailed derivation given in Ref.~\cite{Zou2023prr}. The velocity therefore enters the qubit Hamiltonian on the same footing as the static-field detuning $\varepsilon$, providing a third independent control knob. Unlike the static fields $B_x$ and $B_y$, however, $v$ is a purely dynamical handle: in current-driven racetracks, it is set by the injected current density, on which it depends approximately linearly~\cite{Ryu:2013wh, Yang2015uw, Kim2017natmat, Yang:2021wy, Guan:2021vo, Blasing:2018vb, Yoshimura:2016uk}.}
The ability to manipulate a qubit simply by moving it is a distinctive feature of the domain wall platform. A crucial requirement for the flying qubit is that the domain wall
moves adiabatically: the velocity must be slow enough that the system remains in the qubit subspace and the orbital ground state throughout the transport. This imposes the condition $v \ll \omega_p l_p,\, \omega_0 l_{\text{so}}$, ensuring that the kinetic energy associated with the motion does not excite transitions to higher levels. 
These thresholds are estimated to be of order $10^3$~m/s with realistic parameters~\cite{Zou2023prr}, comfortably above the typical operating regime $v \lesssim 100$~m/s of current-driven domain wall motion in racetrack geometries.

\subsection{DMRG results on discrete spectra and tunneling}
\begin{figure}[!t]
\centering
\includegraphics[width=1\columnwidth]{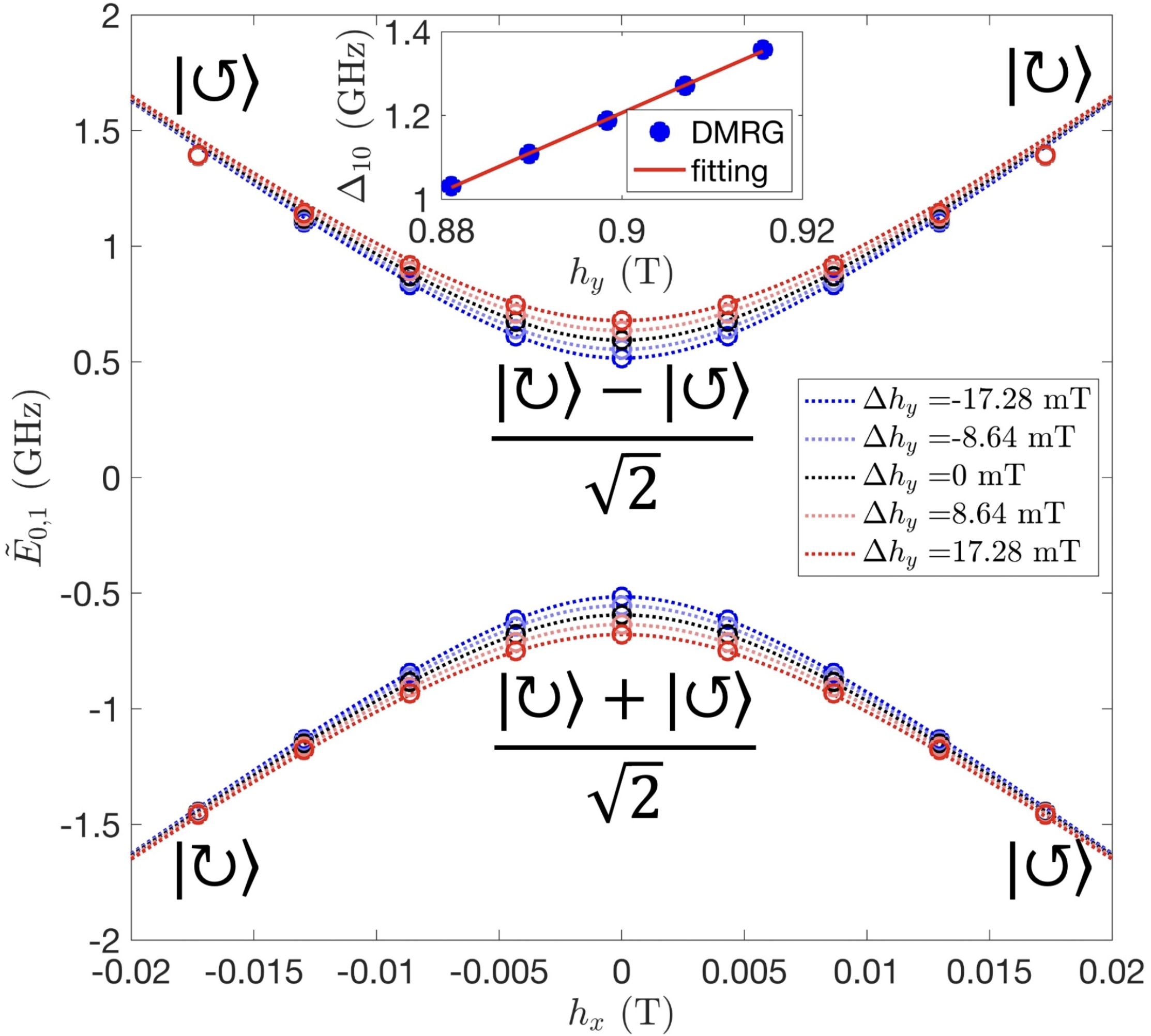}
\caption{\textbf{Domain wall qubit in a quantum spin-$1/2$ chain from DMRG.}
Energy spectrum of the two lowest-lying states of a single domain wall in a ferromagnetic spin-$1/2$ chain as a function of the in-plane field $h_x$, obtained from DMRG simulations (circles) and the effective qubit Hamiltonian Eq.~(\ref{dmrg}) (dotted lines), for several values of the field deviation $\Delta h_y$ from the bias point $h_{y,\mathrm{bias}} = 0.9$~T. At large $|h_x|$ the eigenstates approach definite chirality states $\ket\circlearrowleft$ and $\ket\circlearrowright$, while at $h_x = 0$ they form symmetric and antisymmetric superpositions. The inset shows the qubit splitting $\Delta_{10}$ as a function of $h_y$ near the bias point, exhibiting the characteristic dependence controlled by the tunneling barrier. Parameters: $N = 30$, $J = 25.85$~meV, $K_z = 0.26$~meV, $K_y = 0.1$~meV. Adapted from Ref.~\cite{qu2025density}.}
\label{f3}
\end{figure}

The semiclassical framework developed above treats the domain wall
as a classical soliton whose collective coordinates are subsequently
quantized, an approach that yields analytical control and clear
physical insight. Qu \textit{et al.}~\cite{qu2025density} recently
confirmed that the qubit picture identified above also survives
 in the extreme quantum spin-1/2 limit, by employing the DMRG
method to compute the many-body eigenstates of the full spin
Hamiltonian without any continuum or collective-coordinate
assumptions.
The model is a ferromagnetic spin-$1/2$ chain with the same
symmetry-breaking ingredients as the continuum treatment: isotropic
exchange $J$, easy-axis anisotropy $K_z$, hard-axis anisotropy $K_y$,
and in-plane magnetic fields (denoted $h_x$ and $h_y$ in
Ref.~\cite{qu2025density} and in Fig.~\ref{f3}), while boundary
pinning fields along $\pm z$ enforce a single domain wall in the
ground state. Rather than assuming a domain wall profile and
projecting onto its soft modes, the DMRG treats all spins on equal
footing, providing an unbiased determination of the low-energy
spectrum and its dependence on the applied fields as shown in Fig.~\ref{f3}.

The central finding of the DMRG study is that, for finite hard-axis anisotropy $K_y$, the two lowest-lying eigenstates of the spin chain carry opposite domain wall chiralities and are well isolated from all higher excited states, forming a robust qubit subspace. The chirality of each eigenstate is characterized by the lattice chirality operator
$C_y = \epsilon_{\alpha\beta y}\sum_i S_i^\alpha S_{i+1}^\beta$ with $\alpha, \beta \in \{x, y, z\}$. For  parameters $J = 25.85$~meV, $K_z = 0.26$~meV, $K_y = 0.1$~meV~\cite{qu2025density}, the gap separating this two-dimensional chirality subspace from higher levels exceeds $9.6$~GHz and remains largely insensitive to $h_y$ up to $1$~T, while the qubit splitting $\Delta_{10}$ within the subspace is an order of magnitude smaller. This hierarchy of energy scales is precisely what defines a well-protected qubit: the large gap suppresses leakage, while the tunable splitting governs coherent dynamics. 

Systematically varying the applied fields reveals that $\Delta_{10}$ grows exponentially with $h_y$~\cite{qu2025density}, mirroring the instanton tunneling mechanism of the semiclassical picture, and increases linearly with $h_x$, providing independent control of the detuning. The DMRG energy levels are accurately captured by an effective qubit Hamiltonian:
\( H_\text{DMRG}=\left( \frac{\Delta_{10}}{2} +g_y\Delta h_y \right)\sigma_z-g_xh_x\sigma_x, \label{dmrg}  \)
where $\Delta h_y = h_y - h_{y,\text{bias}}$ is the field deviation
from the bias point $h_{y,\text{bias}} = 0.9$~T, and $g_x \approx
76.2$~GHz/T and $g_y \approx 9.4$~GHz/T are strongly anisotropic
effective $g$-factors that quantify the qubit sensitivity to the
applied field along $x$ and $y$, respectively.

\section{Universal Quantum Computation on Magnetic Racetrack} \label{sec:III}
Having established that the domain wall chirality provides a well-defined qubit degree of freedom, we now turn to the essential ingredients for universal quantum computation: single-qubit control, two-qubit entangling gates, qubit initialization, and readout. 

\subsection{Single Qubit Gate}
The effective spin-orbit coupling of the domain wall qubit translates directly into a powerful control mechanism: moving the domain wall along the racetrack rotates its internal quantum state. This opens several complementary routes to single-qubit gates~\cite{Zou2023prr}. Constant-velocity shuttling produces a rotation around a tunable axis in the $xz$-plane, set by the ratio of tunneling to velocity-induced detuning, while oscillatory velocity modulation at the qubit resonance frequency drives Rabi oscillations with full SU(2) control, the rotation axis being selected by the drive phase. An oscillating magnetic field along $x$ offers an alternative route that couples to the qubit through the large effective $g$-factor $g_x$ identified in the DMRG study~\cite{qu2025density}. Real-time many-body simulations in the spin-$1/2$ limit confirm coherent Rabi oscillations with gate fidelities exceeding $99.9\%$, and the large $g_x \approx 76.2$~GHz/T yields a $\pi$-rotation in approximately $1$~ns using a modest driving field of only a few millitesla~\cite{qu2025density}. 

\subsection{Two-qubit gates}
A universal set of quantum gates requires, in addition to single-qubit rotations, at least one entangling two-qubit operation. In the domain wall platform, such an operation arises naturally from the inter-track interactions between domain walls on neighboring racetracks~\cite{Zou2023prr}. Two domain walls on parallel tracks (or stacked tracks separated by a nonmagnetic spacer) couple via inter-track exchange or dipolar interactions. When projected onto the qubit subspace, this coupling yields an effective interaction of the form 
\begin{equation}
\mathcal{H}^{(2)} = \frac{2N^2 \mathcal{J}_{xx} D}{\sinh D}\, {\sigma}_x^{(1)} \otimes {\sigma}_x^{(2)},
\label{eq:two_qubit_int}
\end{equation}
in the qubit  diagonal basis
which decays exponentially once the inter-wall separation $D$ exceeds the domain wall width $\lambda$. When one domain wall is driven past the other at constant velocity $v$, the accumulated phase generates a controlled-NOT gate (up to single-qubit rotations), with a gate time set by $\lambda/v$. For experimentally accessible interaction strengths $N^2\mathcal{J}_{xx} \sim 50$~MHz and velocities $v \sim 20$~m/s, this yields a two-qubit gate time on the order of $0.2$~ns. The fly-by entangling mechanism is a distinctive feature of the platform: the same current pulse that transports classical bits in racetrack memory now generates quantum entanglement between mobile qubits, with the interaction strength tunable through the inter-track spacing or spacer-layer thickness.

In the DMRG study of a spin ladder geometry~\cite{qu2025density}, where two spin-$1/2$ chains are coupled at a single site, the inter-chain exchange produces the same $\sigma_x \otimes \sigma_x$ interaction structure, with an effective coupling strength that decays exponentially with the separation between the two domain walls~\cite{qu2025density}. Real-time many-body simulations of two mobile domain walls confirm that the fly-by process implements the expected entangling gate with fidelities above $99\%$ and leakage out of the qubit subspace below $10^{-4}$.

In many quantum computing architectures, qubit connectivity is restricted to nearest neighbors, and operations between distant qubits require extensive routing overhead. The domain wall platform offers a natural advantage here: because qubits are inherently mobile, entangling gates between distant domain walls can be performed by shuttling them toward each other until their wave functions overlap, or by dispatching a third domain wall as a flying mediator that sequentially entangles with two stationary qubits at separate locations along the racetrack~\cite{zou2025topological}. These transport-based schemes exploit the intrinsic mobility of the platform without requiring additional coupling hardware. A conceptually distinct alternative is to forgo physical transport altogether and instead let the magnon modes already present in the magnetic wire serve as a quantum bus. Trif and Tserkovnyak~\cite{trif2026cavity} showed that confined magnon modes in short insulating ferromagnetic wires couple to the domain wall through a geometric, Coriolis-type interaction originating from the spin Berry phase. In the dispersive regime, virtual exchange of  magnons generates an effective coupling $J \sim 7$~MHz between two domain wall qubits, yielding a $\sqrt{i\text{SWAP}}$ gate time of approximately $100$~ns. This magnon-mediated mechanism relies entirely on intrinsic magnetic degrees of freedom, requiring no extrinsic photonic cavities or microwave resonators, and offers a complementary route to two-qubit operations with potentially longer-range connectivity than direct exchange coupling can provide.

\subsection{Initialization and readout}
Reliable initialization of the domain wall qubit requires preparing the system in a state with a definite chirality. The simplest approach is to apply a magnetic field $B_x$ along the racetrack, which lifts the chirality degeneracy. At sufficiently low temperatures, the domain wall then relaxes into the ground state with a well-defined chirality.

Readout of the domain wall qubit requires discriminating the two chirality states, and several complementary routes are available. The most direct approach is nanoscale magnetic imaging with NV centers, which can resolve domain-wall textures with spatial resolution on the order of tens of nanometers and, in favorable cases, distinguish chirality through the resulting stray-field pattern~\cite{Song_science_2021,Finco_Natcom_2021,tsukamoto2025observation,tschudin2024imaging}. Another alternative is dispersive readout through a coupled superconducting resonator: in the dispersive regime, the chirality state produces a qubit-state-dependent shift of the cavity resonance, allowing noninvasive detection in close analogy with superconducting circuit QED~\cite{trif2026cavity,tang2025coherent,blais2021circuit}. 
For conducting racetracks, an additional electronically based route is to convert chirality into a local spin-polarization signal, for example via a nearby paramagnetic dot, as proposed theoretically~\cite{Zou2023prr,PhysRevA.57.120}. One may also exploit the chirality-dependent response to current, since domain walls of opposite chirality can move in opposite directions under the same pulse; the resulting displacement can be detected optically or electrically, including with nanoscale Hall-bar arrays that have recently demonstrated~\cite{jeon2024multicore}. This diversity of readout mechanisms, spanning imaging, electronic, and spectroscopic techniques, reflects the rich coupling of domain wall chirality to multiple observable degrees of freedom and increases the likelihood that at least one will prove practical in early experiments.

\section{Material Platforms} \label{sec:platforms}
Realizing the domain wall qubit experimentally requires identifying materials whose intrinsic magnetic properties satisfy the energy hierarchy established in Sec.~\ref{sec:II} and exhibit sufficiently low magnetic damping at millikelvin temperatures. The rapid development of two-dimensional van der Waals magnets, with their strong and tunable anisotropies, atomically clean interfaces, and compatibility with nanoscale fabrication, has produced a class of candidates well suited to these requirements. We first distill the theoretical constraints into concrete material design criteria, then present a detailed quantitative discussion of the most promising candidate CrSBr, and close with a survey of the broader material landscape.

\subsection{Material design criteria} \label{sec:material}
The theoretical framework of Sec.~\ref{sec:II} identifies a clear energy hierarchy, $g\mu_B S_e B,\; K_y < K_z < J$, that any candidate material must satisfy. Translating this hierarchy, together with the coherence and controllability requirements, into experimentally measurable properties yields four essential criteria, summarized in Table~\ref{tab:criteria}.

\emph{Narrow domain walls.} The domain wall width $\lambda = Sa\sqrt{J/K_z}$ controls the number of spins $N$ participating in collective chirality tunneling, which is exponentially sensitive to domain wall size~[Eq.~(\ref{eq:tg})]. 
A small ratio $J/K_z$ is therefore essential, and $N$ ideally should remain in the low hundreds or below for tunneling to reach the experimentally accessible GHz range (see Table~\ref{tab:criteria}).

\emph{Finite hard-axis anisotropy.} A nonzero $K_y$ breaks the $U(1)$ rotational symmetry of the domain wall angle $\Phi$ down to $\mathbb{Z}_2$, generating the double-well potential that defines the
chirality qubit. The ratio $K_y/K_z$ should satisfy $0 < K_y < K_z$ to ensure well-localized chirality states without distorting the domain wall profile.

\emph{Ultralow Gilbert damping at millikelvin temperatures.} A phenomenological estimate of the qubit lifetime gives $T_2 \propto (\alpha\, N\, T)^{-1}$, where $\alpha$ is the Gilbert damping parameter and $T$ is the temperature~\cite{Zou2023prr}. Low damping is therefore as critical as small~$N$. Insulating and semiconducting magnets are systematically preferred because the absence of itinerant electrons removes the dominant spin relaxation channel. At millikelvin temperatures $\alpha$ is expected to be substantially smaller than its room-temperature value~\cite{okada2017, marier2017}, since thermal magnon scattering and phonon-mediated relaxation are both frozen out; however, cryogenic damping remains unmeasured for most magnetic materials, making this one of the most critical open experimental questions for the domain wall qubit program.

\emph{Nanoscale control and experimental access.} The material must support reproducible domain wall nucleation and controlled transport along the racetrack, while the two chirality states must be distinguishable by at least one
experimental probe. Air stability is a further practical consideration, since several otherwise promising van der Waals magnets degrade under ambient conditions~\cite{huang2017layer,burch2018magnetism}, complicating device fabrication. Recent advances in NV center magnetometry~\cite{tschudin2024imaging}  and scanning SQUID microscopy~\cite{zur2023magnetic} have demonstrated nanoscale domain wall imaging in several van der Waals magnets that meet these requirements.

These criteria reframe the classical racetrack design principles for a quantum purpose: the classical program optimizes for thermal stability and fast domain wall motion, whereas the quantum racetrack additionally demands ultralow cryogenic dissipation and a specific anisotropy hierarchy that enables macroscopic quantum tunneling of
the collective coordinate. Finally, while the magnetic order type is not a strict requirement, ferrimagnets are preferred: they offer faster dynamics than ferromagnets, while retaining a net magnetization that enables straightforward field control and detection.

\begin{table}[t]
\caption{\textbf{Material design criteria for domain wall qubits}.
Each row lists the physical requirement, the governing
parameter, and a representative target.}
\label{tab:criteria}
\begin{ruledtabular}
\begin{tabular}{lcc}
Criterion & Parameter & Target \\
\hline
Narrow DW          & $J/K_z$                & $N \lesssim$ few hundred \\
Double well        & $K_y/K_z$              & $\sim 0.1$--$0.5$ \\
Long $T_2$         & $\alpha$ (mK)          & $\lesssim 10^{-5}$ \\
Control \& access  & DW imaging, stability  & Demonstrated \\
\end{tabular}
\end{ruledtabular}
\end{table}

\subsection{CrSBr: a primary candidate} \label{sec:crsbr}

\begin{figure}[!t]
\centering
\includegraphics[width=\columnwidth]{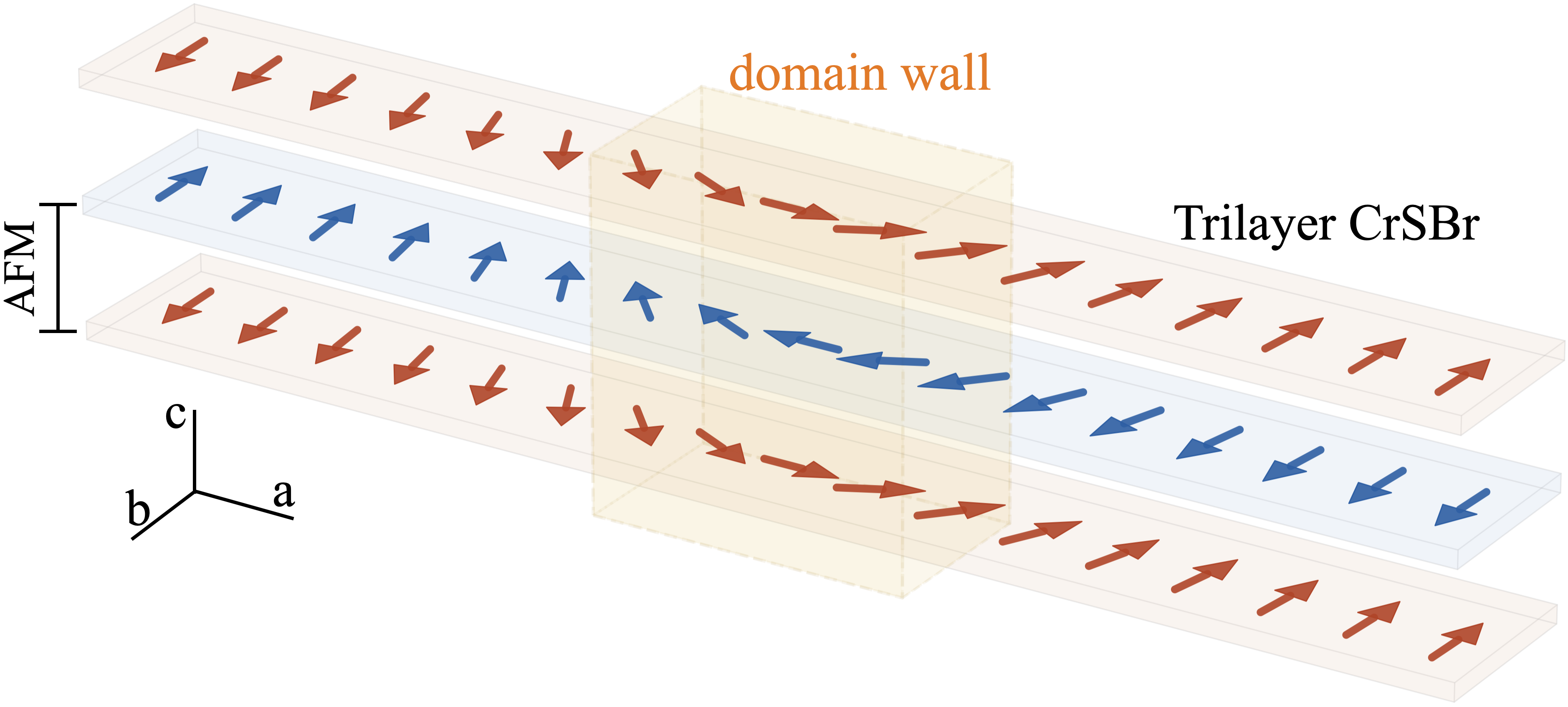}
\caption{%
\textbf{Domain wall in trilayer CrSBr.}
Three CrSBr layers are stacked along the $c$ axis with antiferromagnetic interlayer coupling, forming an effective
ferrimagnet. The strong easy-axis anisotropy along~$b$ and the hard-axis anisotropy along~$c$ stabilize narrow N\'{e}el domain walls in the $ab$ plane, with a characteristic width $\lambda \approx 5.3$\,nm.}
\label{fig:CrSBr}
\end{figure}

Among the growing family of two-dimensional van der Waals magnets~\cite{burch2018magnetism,wang2025magnons}, CrSBr emerges as a compelling candidate that naturally satisfies all the design criteria identified above. Each layer hosts Cr$^{3+}$ ions with spin $S = 3/2$ coupled ferromagnetically within the $ab$ plane by an exchange interaction $J \approx 10$\,meV~\cite{zur2023magnetic,ziebel2024crsbr}, while adjacent layers are antiferromagnetically stacked along the $c$ axis with an interlayer coupling roughly two orders of magnitude weaker. Trilayer films therefore realize an effective ferrimagnetic order with net spin $S_e = 3/2$ per unit cell (see~Fig.~\ref{fig:CrSBr}), combining faster dynamics than a pure ferromagnet with a net magnetization amenable to direct magnetic imaging. 
Equally important from an experimental perspective, CrSBr is chemically stable in air~\cite{ziebel2024crsbr} and supports reproducible domain nucleation in few-layer samples, as confirmed by both nitrogen-vacancy center~\cite{tschudin2024imaging} and scanning SQUID~\cite{zur2023magnetic} magnetometry. 
Besides, its semiconducting character suppresses itinerant-electron relaxation channels and may therefore favor relatively low magnetic damping~\cite{ziebel2024crsbr}.

CrSBr possesses strong magnetocrystalline anisotropies: a pronounced easy axis along $b$ with $K_b \approx 0.1$\,meV and a hard axis along $c$ with $K_c \approx 0.02$\,meV~\cite{zur2023magnetic}, both stable from bulk to the monolayer limit. In the notation of Sec.~\ref{sec:II}, $K_b$ plays the role of the easy-axis anisotropy $K_z$ that confines the domain wall as sketched in~Fig.~\ref{fig:CrSBr}, while $K_c$ plays the role of the hard-axis anisotropy $K_y$ that generates the chirality double-well potential. The strong easy-axis anisotropy produces a domain wall width of only $\lambda = Sa\sqrt{J/K_b} \approx 5.3$\,nm (with lattice constant $a = 0.35$\,nm along the wire direction), keeping the number of spins participating in collective chirality tunneling in the range $N \sim 15$ (single chain) to several hundred
(trilayer strip of finite cross section).

The dimensionless field $b_i = \pi\hbar S_e \gamma B_i / 4K_c$ that governs the potential~[Eq.~(\ref{eq:potential})] converts to physical units via $B_i = b_i \times 150\;\text{mT}$, where $\gamma/2\pi \approx 28$\,GHz/T is the gyromagnetic ratio. The applied magnetic field provides two independent control knobs
on the qubit Hamiltonian~[Eq.~(\ref{eq:qubit_H})]: $B_y$ (along the hard $c$-axis, with operating values $\sim$60 to 90\,mT) controls the tunneling barrier and hence the tunneling amplitude~$t_g$, while $B_x$ (also in the mT range) along the  $a$-axis tunes the chirality detuning~$\varepsilon$. Both lie comfortably within the reach of standard vector-magnet cryostats or on-chip micromagnet arrays.

\begin{figure}[!t]
\centering
\includegraphics[width=\columnwidth]{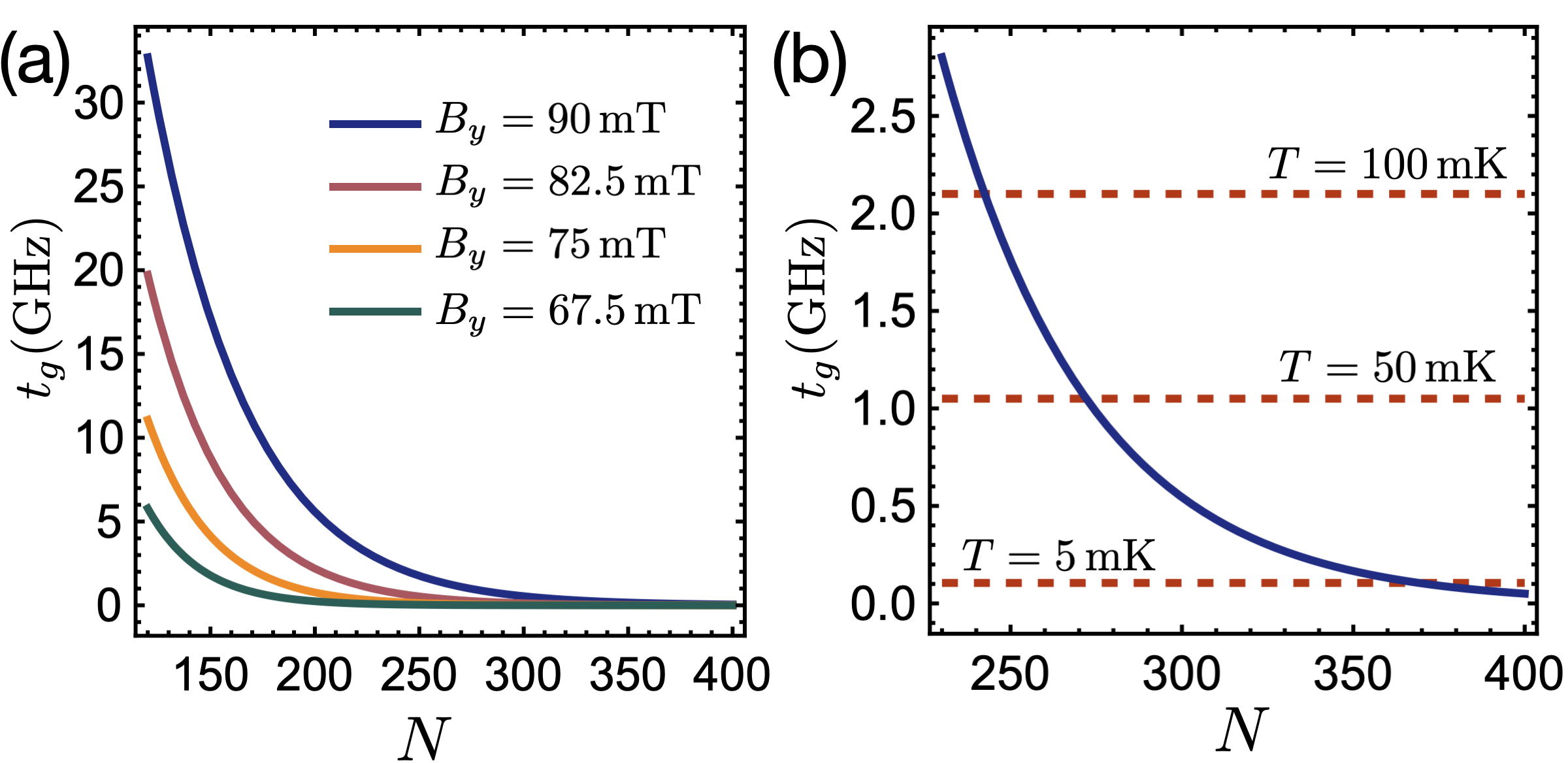}
\caption{%
\textbf{Chirality tunneling rate for CrSBr domain wall qubits.}
(a)~Tunneling splitting $t_g$ as a function of the number of
spins $N$ within the domain wall for several values of the
applied magnetic field $B_y$ along the hard $c$~axis.
The tunneling rate decreases exponentially with $N$ and
increases with $B_y$, which suppresses the potential barrier
between the two chirality states.
(b)~Tunneling splitting at $B_y = 90$\,mT compared with the
thermal energy scale $k_BT/h$ (dashed lines) for three
operating temperatures.
Quantum coherence requires $t_g > k_BT/h$:
at $T = 5$\,mK this condition is satisfied for domain walls
containing up to $N \approx 400$ spins, while at
$T = 100$\,mK the bound tightens to $N \lesssim 250$.
All curves are evaluated using CrSBr parameters
($J = 10$\,meV, $K_b = 0.1$\,meV, $K_c = 0.02$\,meV).
}
\label{fig:tunneling}
\end{figure}

Figure~\ref{fig:tunneling}(a) displays the tunneling splitting $t_g$
as a function of the number of spins $N$ for several values of $B_y$.
At $B_y = 90$\,mT ($b_y = 0.6$), tunneling rates in the GHz range
persist up to $N \approx 200$, while at $B_y = 67.5$\,mT
($b_y = 0.45$) the same rate requires $N \lesssim 150$.
Throughout this regime, the level spacing
$\hbar\omega_0 = 2\sqrt{2JK_c(1-b_y^2)}$,
which sets the gap protecting the qubit subspace from higher
excitations, exceeds the tunneling splitting by more than an order
of magnitude, ensuring a well-isolated two-level system.
Temperature imposes a complementary constraint~[Fig.~\ref{fig:tunneling}(b)]:
the tunneling rate must exceed the thermal energy scale $k_BT/h$
for quantum coherence to prevail over thermal fluctuations.
At $T = 5$\,mK this condition is met for $N$ up to approximately 400,
while at $T = 100$\,mK the bound tightens to $N \lesssim 250$ when $B_y=90$ mT.
Millikelvin operation, already standard for superconducting and
spin qubit experiments, is therefore essential.

The qubit coherence time can be estimated phenomenologically from the Gilbert damping. In the dephasing-dominated regime, where the detuning $\varepsilon$ exceeds the tunneling amplitude $t_g$, this yields~\cite{Zou2023prr}
\(
T_2 = \frac{\alpha^{-1}}{2\pi S\, N\, (k_BT/h)},
\label{eq:T2}
\)
where $\alpha$ is the Gilbert damping parameter. The inverse scaling with both $N$ and temperature $T$ reinforces
the need for narrow domain walls and millikelvin operation. For $N = 100$ and $T = 10$\,mK, this gives $T_2 = \alpha^{-1} \times 5 \times 10^{-6}\,\mu$s. Two experimental benchmarks provide guidance for the expected dissipation at cryogenic temperatures. 
First, the van der Waals ferromagnet Cr$_2$Ge$_2$Te$_6$ exhibits damping comparable to yttrium iron garnet 
(YIG, $\alpha \sim 10^{-4}$--$10^{-5}$) in the 20 to 50\,K range~\cite{xu2023electrical}. 
Second, magnon lifetimes in YIG reach $\sim 20\,\mu$s at 30\,mK~\cite{serha2025ultra}, indicating a 
strong suppression of dissipation as thermal scattering processes freeze out. 
CrSBr, as a semiconductor, is expected to suppress itinerant-electron relaxation channels that dominate 
damping in metallic magnets, and may therefore reach comparably low values of $\alpha$. 
{If the cryogenic damping reaches $\alpha \sim 10^{-5}$–$10^{-6}$, the corresponding coherence time would be of order $T_2 \sim 0.5$–$5\,\mu$s; for a more conservative $\alpha \sim 10^{-4}$ it would fall to $\sim 50\,$ns. A quantitative value therefore awaits a direct measurement of the millikelvin damping (Sec.~\ref{subsection:dissipation}).}
Combined with single-qubit gate times of $\sim$1\,ns (Sec.~\ref{sec:III}), these estimates yield a quality factor $Q = T_2/t_{\rm gate} \sim 0.5 \times 10^3$ to $0.5 \times 10^4$ {in the favorable case, and still $Q \approx 50$ for the conservative $T_2 \sim 50\,$ns, so that many operations fit within one coherence time}. For comparison, early superconducting~\cite{nakamura1999coherent} and semiconductor spin qubit~\cite{Petta2005science} devices often operated with similarly modest quality factors before sustained engineering improved coherence by orders of magnitude. If the damping proves favorable, domain wall qubits could therefore enter the field from a competitive starting point.

The semiconducting character of CrSBr further expands the available control space. Electrostatic gating can, in principle, influence magnetic interactions and anisotropies through carrier modulation, offering a potential route toward electrical control that complements, and may  replace, local magnetic field tuning. Strain applied via piezoelectric substrates offers a complementary static handle on the anisotropy ratio $K_c/K_b$.
Interfacing CrSBr with heavy-metal layers could introduce a Dzyaloshinskii-Moriya interaction that serves as an intrinsic chirality detuning, potentially replacing the applied field $B_x$. While these additional mechanisms remain to be quantified experimentally, they demonstrate that the platform supports a rich landscape of independent control knobs, offering a degree of in-situ tunability competitive with the most established qubit platforms.

\subsection{Broader material landscape}
While CrSBr provides a particularly favorable platform, the design principles outlined in Sec.~\ref{sec:material} are general and can guide the exploration of a broader class of magnetic materials. Within the rapidly expanding family of two-dimensional van der Waals magnets~\cite{burch2018magnetism}, several alternative systems may offer complementary advantages.

Fe$_3$GeTe$_2$ is a metallic itinerant ferromagnet with gate-tunable Curie temperature and strong perpendicular anisotropy in thin films~\cite{deng2018gate,fei2018two}, making it highly attractive for spintronic applications. Current-driven domain wall motion has also been demonstrated in this system~\cite{zhang2024current}, highlighting its promise for racetrack-style architectures in which domain walls can be generated, transported, and manipulated on demand. Its metallic character introduces additional damping channels associated with conduction electrons, but also enables direct electrical control schemes not readily available in insulating systems, suggesting a potential trade-off between tunability and dissipation.

CrI$_3$ is an insulating magnet with strong Ising-type anisotropy~\cite{huang2017layer}, a feature favorable for suppressing dissipation. While its anisotropy is predominantly uniaxial, additional anisotropy components may be engineered through strain, stacking, or proximity effects~\cite{webster2018strain,vishkayi2020strain,zhang2021emergent}, providing possible routes toward realizing the effective double-well potential required for domain wall qubits.

Cr$_2$Ge$_2$Te$_6$ has demonstrated exceptionally low Gilbert damping, comparable to yttrium iron garnet in the 20 to 50\,K range~\cite{xu2023electrical}. Its relatively weak magnetic anisotropy~\cite{suzuki2022magnetic,zeisner2019magnetic} leads to wider domain walls and suppressed tunneling rates. Combining such low-damping materials with stronger-anisotropy systems in van der Waals heterostructures~\cite{zhong2017van} offers a promising route to optimize both coherence and confinement.

Beyond two-dimensional materials, rare-earth iron garnets, such as YIG, set the benchmark for ultralow magnetic dissipation, with $\alpha \sim 10^{-5}$ in thin films and even lower effective dissipation at millikelvin temperatures~\cite{serha2025ultra}. Although their domain walls are typically wide, advances in nanostructuring and strain engineering may enable control over their effective size and dynamics~\cite{mendil2019magnetic,al2025strain}, opening possible connections to quantum regimes.

Looking forward, the continued discovery of new van der Waals magnets and rapid progress in materials engineering together reveal a rich and largely unexplored design space for future quantum platforms.

\section{Key experiments}\label{sec:experiments}
The theoretical framework and a promising material platform are now in place; what remains is to demonstrate that domain wall chirality behaves quantum mechanically. We organize the path forward into four steps: measuring cryogenic magnetic dissipation, resolving the chirality two-level system spectroscopically, demonstrating coherent qubit control and readout, and exploiting the intrinsic domain wall mobility for transport and entanglement. Many of the required techniques already exist in the racetrack, magnon spintronics, and circuit QED communities; the principal challenge is to combine them at millikelvin temperatures and
at the single domain wall level.

\subsection{Cryogenic magnetic dissipation} \label{subsection:dissipation}
The coherence analysis of Sec.~\ref{sec:crsbr} identifies magnetic damping at millikelvin temperatures as the parameter that  directly controls the viability of the domain wall qubit. For CrSBr, however, the relevant low-temperature damping remains insufficiently characterized, especially in the millikelvin regime. 

Several experimental approaches could be pursued to close this gap. First, broadband microwave resonance offers the most direct probe of low-temperature magnetic damping. In CrSBr, magnetic resonance has already been measured as a function of frequency, field orientation, and temperature, establishing microwave spectroscopy as a natural platform for linewidth-based damping extraction~\cite{cham2022anisotropic}. An important next step is to extend these measurements to micropatterned flakes in a dilution refrigerator, enabling linewidth measurements down to the millikelvin regime. Second, cryogenic Brillouin light scattering~\cite{sandweg2010wide} can provide mode-selective and wave-vector-sensitive access to magnon spectra and linewidths, complementing the zone-center information obtained from conventional microwave measurements. Third, time-resolved pump-probe measurements can follow the decay of magnetic dynamics directly in the time domain~\cite{kirilyuk2010ultrafast,hendriks2024electric}, and may uncover multichannel or nonexponential relaxation that would remain hidden in a simple linewidth analysis.

\subsection{Spectroscopic identification of chirality levels}

Another important step is to identify the chirality two-level system as a discrete spectroscopic feature. As illustrated in Fig.~\ref{fig:tunneling}, the two lowest states of a single confined domain wall form a qubit with a splitting in the GHz range that is tunable by the applied field $B_y$. These states are separated from higher excitations by a substantially larger gap, helping to isolate the low-energy two-level subspace.

A key prediction is the strongly anisotropic field dependence of the splitting: it depends exponentially on $B_y$, which controls the tunneling barrier, and linearly on $B_x$, which controls the detuning. This combination provides a distinctive experimental fingerprint that should help distinguish the chirality qubit from other low-energy magnetic modes.

The most direct probe is microwave spectroscopy~\cite{cham2022anisotropic} on a CrSBr nanowire containing a single pinned domain wall, searching for a resonance in the GHz range whose frequency follows the predicted field dependence. A more sensitive alternative is to couple the domain wall to a high-$Q$ superconducting coplanar-waveguide resonator~\cite{tang2025coherent}. In the dispersive regime, the chirality state can then produce a state-dependent cavity frequency shift, closely analogous to dispersive readout in circuit QED~\cite{blais2021circuit}. Either approach would provide access to the effective couplings $g_x$ and $g_y$, enabling a quantitative comparison between experiment and theory. Observation of a resonance with the predicted anisotropic field dependence would constitute direct evidence that domain wall chirality forms a quantum two-level system.

\subsection{Single domain wall qubit operation}
Once the two-level system has been identified spectroscopically, the next step is to demonstrate initialization and coherent control of a single confined domain wall. The wall must first be localized at a designated position. Geometric constrictions patterned into the nanowire provide a natural route to domain-wall trapping~\cite{im2009direct,yuan2014domain}, while gate-controlled pinning potentials, potentially enabled by the semiconducting character of CrSBr, offer a promising longer-term alternative~\cite{jo2024nonvolatile}. The confinement should be strong enough that the positional uncertainty $l_p$ remains comparable to or smaller than the spin-orbit length $l_{\rm so}$, so that the tunneling amplitude is not strongly suppressed.
Initialization into a definite chirality state can be achieved by passive thermalization under an applied field $B_x$, which lifts the chirality degeneracy. An alternative route is magnon-assisted sideband cooling to the ground state~\cite{trif2026cavity}, analogous to resolved-sideband protocols in trapped-ion and cavity-based platforms.

Coherent control can be implemented by driving the chirality transition with an oscillating magnetic field along $x$. Owing to the large effective $g$-factor in this direction, a resonant ac magnetic field of only a few millitesla can flip the qubit on the nanosecond timescale. Observation of Rabi oscillations and their decay would establish coherent control and provide a first measure of the coherence time. Subsequent Ramsey and echo protocols could then distinguish whether decoherence is dominated by the intrinsic Gilbert damping or by extrinsic sources, such as electrical noise in the device environment or fluctuations from the nuclear-spin bath.

%For readout, NV center magnetometry can, in principle, distinguish the stray-field patterns of opposite chiralities and is a promising probe for early experiments~\cite{tsukamoto2025observation,Song_science_2021, Finco_Natcom_2021}. Another alternative is dispersive readout through a coupled superconducting resonator~\cite{tang2025coherent}, which offers high bandwidth and direct compatibility with circuit-QED architectures~\cite{Blasi2021rmp}. The central challenge is sensitivity: the magnetic signal of a single nanoscale domain wall is weak, and reliable chirality discrimination requires spatial resolution comparable to, or better than, the domain-wall width $\lambda$. Achieving chirality-resolved readout by either route would complete the basic qubit toolbox.

\subsection{Coherent transport}

A defining advantage of the domain-wall platform is its intrinsic mobility: in principle, the same mechanisms used to shuttle classical bits in racetrack memory can also transport a quantum state from one location to another. Demonstrating that such motion preserves quantum coherence would establish a functionality that is unavailable in most stationary solid-state qubit platforms.

A central challenge is Joule heating. The spin-polarized currents commonly used to drive domain walls dissipate energy in the nanowire and can raise the local temperature well above the dilution-refrigerator base, thereby degrading quantum coherence. One possible strategy is to decouple the drive from the quantum track by placing a classical racetrack, for example a metallic magnetic layer with robust domain walls, adjacent to the qubit track. Motion in the classical layer could then drag the quantum domain wall through interlayer dipolar or exchange coupling, while keeping the quantum track current-free. Current-free alternatives are also worth exploring, including magnetic field-driven transport and propulsion by surface acoustic waves~\cite{rivelles2025moving}.

\section{Outlook}
The domain wall qubit concept opens research directions that extend well beyond the specific platform considered here. In this closing section, we highlight several particularly promising themes: extending the qubit paradigm to other topological magnetic textures and electrically controlled implementations, embedding domain wall qubits in hybrid quantum architectures that exploit their intrinsic mobility, and addressing open theoretical questions whose resolution would sharpen quantitative predictions and deepen our understanding of macroscopic quantum coherence in magnetic solitons. Progress along these directions would also advance the broader fields of quantum magnonics and spintronics.

\subsection{Qubits from topological spin textures}

The domain wall qubit exploits a single internal degree of freedom of one particular topological soliton, but the
underlying principle is far more general: topological magnetic textures generically possess collective coordinates whose quantization in engineered double-well potentials yields two-level systems.
This observation opens a family of soliton-based qubits, each with distinct properties that may prove advantageous for different tasks.

Magnetic skyrmions, whose internal helicity angle can tunnel between two potential minima through the same instanton mechanism as the domain wall chirality, have also been proposed as qubit candidates~\cite{christina_prl_2021,psaroudaki2023skyrmion}. But the use of skyrmions as mobile qubits remains largely unexplored. This possibility is especially intriguing because skyrmions respond to current in ways that differ qualitatively from domain walls, most notably through their transverse dynamics and the skyrmion Hall effect, which could open new routes for qubit transport and control.
Magnetic vortices, which form spontaneously in confined geometries such as nanodisks~\cite{zou2025tunable}, offer a complementary realization: the vortex helicity can undergo macroscopic quantum tunneling between degenerate configurations, and the resulting two-level system can be controlled through tunable detuning and external fields, with neighboring vortex qubits coupled via exchange interactions or magnon-mediated processes.
At the frontier of topological complexity, three-dimensional textures such as magnetic hopfions~\cite{liu2018binding}, classified by the Hopf invariant, possess richer internal structure that could in principle encode multi-level Hilbert spaces within a single soliton. At present, however, the quantum dynamics of such higher-dimensional solitons remain largely unexplored.

A common thread across all these textures is the question of efficient control. If the host material is multiferroic, with coupled magnetic and electric order parameters~\cite{li2023energy,fert2024electrical}, electric fields could directly modulate the collective-coordinate potential, enabling all-electrical qubit manipulation without local magnetic field sources.
This prospect is especially appealing for scaling, as gate electrodes offer nanometer spatial precision, and it applies
equally to domain wall, skyrmion, vortex, and hopfion qubits. 
%More broadly, the theoretical tools developed for the domain wall qubit, collective-coordinate quantization, and instanton calculations, transfer directly to each of these systems, establishing a unified framework for soliton-based quantum information processing.

\subsection{Hybrid architectures and quantum buses}

\begin{figure}[!t]
\centering
\includegraphics[width=\columnwidth]{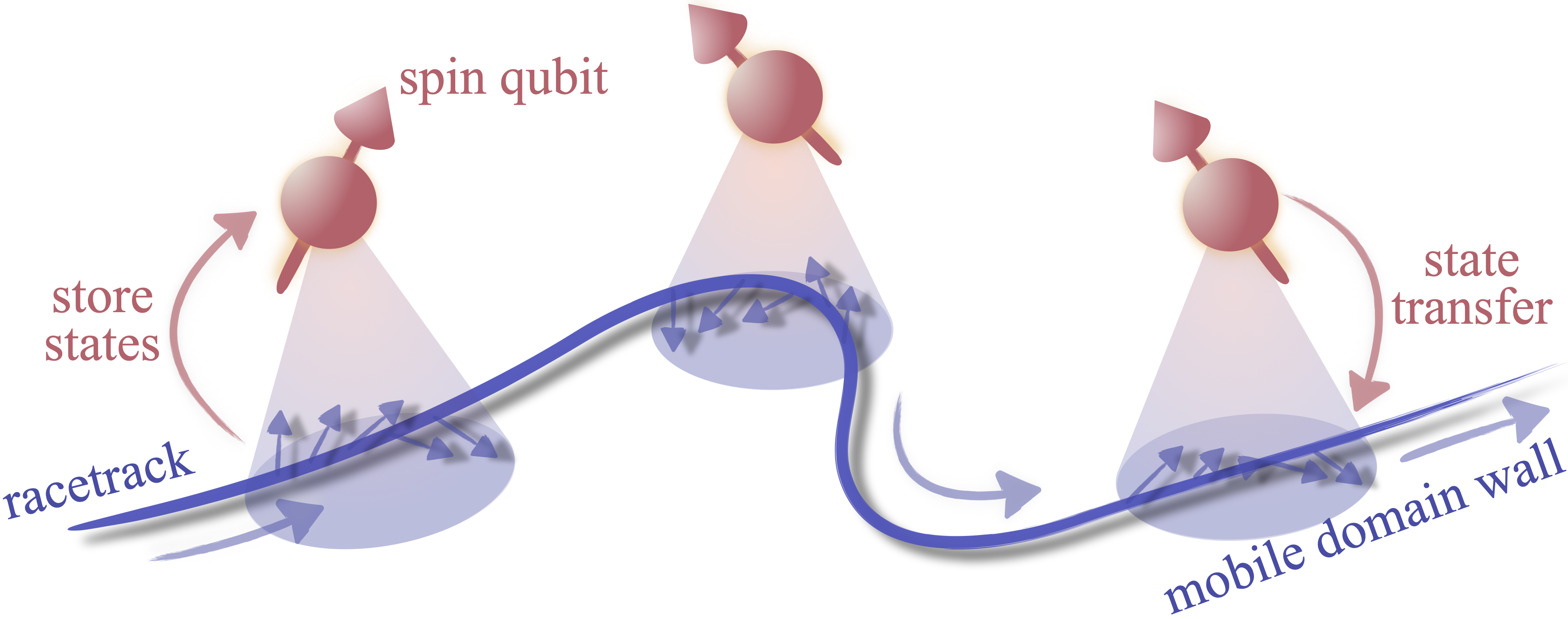}
\caption{%
\textbf{Concept of a racetrack-based quantum bus.}
A mobile domain wall moves along a magnetic racetrack and couples sequentially to distant qubits, allowing quantum states to be stored, transported, and transferred between separated nodes. This illustrates how domain wall mobility can provide a solid-state route to nonlocal quantum connectivity. Adapted from Ref.~\cite{zou2025topological}.
}
\label{fig6}
\end{figure}

An important future direction is to embed domain wall qubits in hybrid quantum architectures, where different physical platforms contribute complementary strengths. In such settings, domain walls can serve not only as qubits, but also as mobile, magnetically active quantum links that interface naturally with actively pursued platforms such as superconducting and spin qubits.

One promising direction is to couple domain wall qubits to superconducting circuits. Although such an interface has not yet been developed specifically for domain walls, the broader quantum-magnonics literature has already established coherent coupling between superconducting qubits and magnetic excitations through microwave photons and magnetic near fields, making analogous hybrid architectures plausible here~\cite{tabuchi2015coherent,lachance2020entanglement}. Domain wall qubits could also couple to spin qubits in quantum dots, either through the dipolar stray field of the wall or, at sufficiently short distances, through exchange interactions. A domain wall qubit could be physically shuttled between different processing nodes on a chip, interacting sequentially with multiple modules~\cite{Zou2023prr,zou2025topological}. In this way, a racetrack bus could mediate entanglement or quantum state transfer between otherwise weakly coupled nodes, playing a role analogous to a solid-state flying qubit, as shown in Fig.~\ref{fig6}. Such an approach directly addresses the connectivity bottleneck of many existing solid-state architectures and may be especially attractive for modular processors or for protocols requiring nonlocal operations.

\subsection{Open theoretical questions}

Several theoretical questions remain unresolved. A central challenge is to develop a microscopic theory of decoherence. Throughout this Perspective, coherence estimates have been based on a phenomenological connection between dephasing and Gilbert damping, guided by the fluctuation-dissipation theorem. A more complete treatment should identify the dominant noise sources, including thermal magnons, phonons, substrate-induced charge noise, and the nuclear-spin bath of the constituent atoms. Such an analysis would clarify which decoherence channels are intrinsic to the material and which could instead be reduced through materials engineering or device design, thereby helping to inform possible error-mitigation and noise-correction strategies.

Another question concerns the robustness of flying domain wall qubits under realistic conditions. While mobility is a defining advantage of the racetrack platform, transport inevitably exposes the qubit to spatial inhomogeneities, disorder-induced pinning, and time-dependent perturbations that may induce dephasing or leakage out of the chirality subspace. A quantitative theory of coherent domain wall transport in the presence of such imperfections remains largely undeveloped. Developing such a framework will be essential for assessing the fidelity of transport-based operations and motion-enabled gate protocols.

These considerations naturally extend to the level of quantum architecture and error correction. Because domain wall qubits can be physically rearranged, the platform offers a form of dynamical, nonlocal connectivity that is difficult to realize in architectures with fixed qubit locations. This raises the question of which code families can best exploit such reconfigurability. Quantum LDPC codes are a particularly intriguing possibility, since their nonlocal stabilizer structure is costly to implement on strictly nearest-neighbor hardware. In a racetrack architecture, by contrast, the cost of a nonlocal operation could in principle be set by physical transport rather than by a sequence of SWAP gates. Developing error-correcting schemes tailored to this geometry, and understanding the tradeoff between transport-induced errors and the connectivity advantage provided by reconfigurability, therefore remains an open problem with potentially broad implications for quantum architecture design.

\subsection{Toward a Quantum Racetrack}
In this Perspective, we have outlined a possible path from  classical racetrack memory proposed in Ref.~\cite{Parkin190} to a quantum racetrack built from the same underlying objects. The classical racetrack established that trains of magnetic domain walls can reliably store and transport classical bits along nanowires driven by spin-polarized currents. The quantum racetrack asks whether these same topological solitons, now engineered at the nanoscale and operated at millikelvin temperatures, can also store, transport, and process quantum information. The framework developed here, combining semiclassical quantization, DMRG results in the fully quantum spin-$1/2$ limit, and the quantitatively promising material platform CrSBr, suggests that this possibility deserves to be taken seriously.

{This path is also a scalable one. Because the qubits are mobile, two-qubit gates can be realized by transporting walls into proximity or by a domain wall acting as a flying quantum bus, so the same current pulses that shift classical bits in racetrack memory also entangle distant qubits, directly addressing the connectivity bottleneck of most static solid-state platforms. A scalable device would then comprise arrays of such racetracks in a material meeting the criteria of Sec.~\ref{sec:material}, with CrSBr a concrete candidate; the experiments of Sec.~\ref{sec:experiments} chart a path from a single domain wall qubit toward a multi-qubit quantum racetrack.}

More broadly, the domain wall qubit embodies the idea that topological solitons in magnets, long central to classical information technology, can be promoted into the quantum regime. If realized, this vision would do more than add another platform to the quantum computing landscape. It would establish magnetic domain walls as a new setting in which to study macroscopic quantum coherence, the quantum-classical boundary of collective topological objects, and the interplay between topology, motion, and quantum information. The path forward will require close collaboration across magnetism, quantum information, and materials science. The potential reward, however, is substantial: a mobile qubit that naturally unifies information storage, transport, and processing within a single physical object, and a new route toward scalable quantum architectures built from magnetic solitons.

\subsection{Acknowledgement}
This work was
supported as a part of NCCR SPIN, a National Centre of Competence in Research, funded by the Swiss National Science Foundation (grant number 225153). D.L. acknowledges the Deanship of Research and the Quantum Center for the support received under Grant no.
CUP25102 and no. INQC2600, respectively.

\subsection{Conflicts of Interest}
The authors declare no conflicts of interest.

\subsection{Data Availability Statement}
Data sharing not applicable to this article as no datasets were generated or analysed during the current study.

%\bibliography{/Users/zouji/Dropbox/!Paper/Paper.bib}
%apsrev4-2.bst 2019-01-14 (MD) hand-edited version of apsrev4-1.bst
%Control: key (0)
%Control: author (8) initials jnrlst
%Control: editor formatted (1) identically to author
%Control: production of article title (0) allowed
%Control: page (0) single
%Control: year (1) truncated
%Control: production of eprint (0) enabled
%

\end{document}